\documentclass[11pt]{article}
\usepackage{graphicx}       
\newcommand{\re}{\mathrm{e}} 
\newcommand{\hp}{\mathrm{h}}
\newcommand{\ri}{\mathrm{i}}

\newcommand{\ru}{\mathrm{u}}
\newcommand{\rv}{\mathrm{v}}
\newcommand{\intk}{ \int_{- \pi / \ell}^{\pi/ \ell} dk}
\newcommand{\nz}{\mathbb{Z}}
\newcommand{\kb}{\mathrm{k}_{\mathrm B}}
\RequirePackage{amssymb}
\usepackage{geometry} 
\begin{document}

\begin{center}
{\bf Quantum Distributions for the Electromagnetic Field} \\[2cm] 
M. Grigorescu \\[3cm]  
\end{center}
\noindent
$\underline{~~~~~~~~~~~~~~~~~~~~~~~~~~~~~~~~~~~~~~~~~~~~~~~~~~~~~~~~
~~~~~~~~~~~~~~~~~~~~~~~~~~~~~~~~~~~~~~~~~~~~~~~~~~~~~~~~~~}$ \\[.3cm]
The coherence properties of the classical waves are discussed in terms of the Cauchy 
problem for the wave equation, and of a discrete representation by an ensemble of
Hamiltonian systems. Wave quanta are related to specific "action fields", and phase-space
distributions of phonons and photons are obtained by Wigner transform. For photons in 
a thermal environment, the proposed Wigner function evolves towards the Planck equilibrium
distribution. It is shown that  the free electromagnetic field can also be found in states of
definite helicity,  described  by a complex vector potential. 
\\
$\underline{~~~~~~~~~~~~~~~~~~~~~~~~~~~~~~~~~~~~~~~~~~~~~~~~~~~~~~~~~~~~~
~~~~~~~~~~~~~~~~~~~~~~~~~~~~~~~~~~~~~~~~~~~~~~~~~~~~~}$ \\

\newpage

\section{Introduction} 
The great discoveries of X-rays,  with particle-like properties  (R\"ontgen, 1895), and 
of the energy quanta, explaining the spectral distribution of thermal radiation (Planck, 1900), 
soon after the discovery of radio waves (Hertz, 1887), have re-opened the old debate on the 
particle-wave nature of light, with partial solutions, and updates, until today. Now QED is 
a well established theory \cite{bd}, and an example of accuracy for its predictions at the 
atomic level, but still, there is no clear relationship between some of its basic 
elements, like the photons, and the classical theory \cite{jdj}. A detailed discussion of 
optical phenomena such as diffraction and refraction of light, using the Feynman's path 
integrals, can be found in \cite{jhf}, while various aspects concerning a photon wave function 
defined by a complex linear combination of the classical fields ${\bf E}, {\bf B}$ are presented in \cite{bb, smra}. Though, these fields are not canonical variables \cite{whwp}, and the outcome provides energy distributions\footnote{For the electromagnetic field in  a medium there is also the 
open problem of the energy-momentum tensor $[T_{ik}]$, usually presented in the Minkowski (standard), or Abraham (symmetric) form \cite{ne}.}  rather than probability density. 
\\ \indent
This work is an attempt to describe the wave quanta (phonons and photons) by specific 
"action waves", and distributions on a phase-space with granular structure. However, instead of considering the limit of vanishing rest mass\footnote{In this limit  the Compton wavelength, which sets the scale for space discretization, becomes infinite. Besides, wave quanta may have $E/p  \ll c$.}  in the relativistic Schr\"odinger equation for massive particles \cite{rpw}, the starting point here is the real, observable field. The basic elements of the present approach are introduced in Section 2, using the simple example of a 1-dimensional lattice of coupled harmonic oscillators. In Section 3, after a brief recall of the geometric framework behind the Maxwell equations, the photon wave function is defined, and its phase-space representation, obtained by the Wigner transform, is discussed both at zero and finite temperature. Concluding remarks are summarized in Section 4.               

\section{The phonon wave modes and phase-space for a 1d lattice }
Let us consider the phase space $M= \nz \times \mathbb{R}^2$, parameterized by the 
real canonical variables $\{(\ru_n, \rv_n) \in \mathbb{R}^2, n \in   \nz \}$, with the 
Poisson brackets $\{\ru_n, \rv_{n'} \} = \delta_{nn'}$,  $\{\ru_n, \ru_{n'} \} = 0$, 
$\{\rv_n,\rv_{n'} \} = 0$. For a system of coupled harmonic oscillators, described by the 
Hamiltonian  \cite{agar} 
\begin{equation}
H= \sum_{n \in \nz} [ \frac{\rv_n^2}{2m} + \frac{m \omega_0^2}{2} \ru_n^2   
 +   \frac{m \kappa}{2} (\ru_n- \ru_{n-1})^2 ]~~,~~ \kappa >0~~, \label{ham1} 
\end{equation}
the equations of motion $\dot{\ru}_n = \{\ru_n, H \}$,  $\dot{\rv}_n = \{\rv_n, H \}$, lead to
\begin{equation}
\ddot{\ru}_n = - \omega_0^2 \ru_n + \kappa (\ru_{n+1} + \ru_{n-1} - 2 \ru_n)~~,~~n \in \nz ~~. 
 \label{ddu}
\end{equation} 
With respect to a unit of length $\ell \in \mathbb{R}_+$, (the lattice constant), one can 
define a real index $x= n \ell \in \ell \nz$, and new coordinates $ \ru_{n \ell} \equiv
\ru_n / \sqrt{\ell}$,  $ \rv_{n \ell} \equiv \rv_n / \sqrt{\ell}$, such that $\{\ru_{n \ell}, 
\rv_{n' \ell} \} = \delta_{nn'} / \ell$. Presuming that in the limit $ \ell \rightarrow 
\ell_0 > 0$, $x$ can be considered as a continuous variable,  then  $\{\ru_x, \rv_{x'} \} \rightarrow \delta(x-x')$, $\ell \sum_n \rightarrow \int dx$, and    
\begin{equation}
H \rightarrow  \int dx~ [ \frac{\rv_x^2}{2m} + \frac{m \omega_0^2}{2} \ru_x^2   
 +   \frac{m v^2}{2} (\partial_x \ru)^2 ]~~, 
\end{equation}
where $v= \ell_0 \sqrt{\kappa}$, $\partial_x \equiv \partial / \partial x$. In the continuous 
limit (\ref{ddu}) becomes  $ \partial^2_t{\ru} + \omega_0^2 \ru= v^2 \partial^2_x \ru$,
as $(\ru_{x+\ell} + \ru_{x-\ell} - 2 \ru_x)/ \ell^2$ is the finite differences expression of 
$\partial^2_x \ru$. When $\omega_0=0$ this reduces to the plane wave equation without sources,  $\partial^2_t{\ru}= v^2 \partial^2_x \ru$, having the known solution 
\begin{equation} 
\ru(t) = \mathcal{E}_t  \tilde{*} \dot{\ru}^0 + \partial_t \mathcal{E}_t \tilde{*} \ru^0~~,  
\label{spwe}     
\end{equation}
where $\mathcal{E}_t$ is the fundamental solution of $(\partial^2_t- v^2 \partial^2_x ) 
\mathcal{E}_t = \delta(x) \delta(t)$,  $\tilde{*}$ denotes the convolution product, and $\dot{\ru}^0 \equiv  \partial_t \ru \vert_{t=0}$, $\ru^0 \equiv \ru \vert_{t=0}$, are the initial conditions of the Cauchy problem. An alternative formulation can be given by a partial 
decoupling of the oscillators using the complex coordinates 
\begin{equation}   
\ru'_\alpha = \frac{1}{\sqrt{2 \pi}} \sum_{n \in \mathbb{Z}} \re^{\ri \alpha n} \ru_n~~,~~
\rv'_\alpha = \frac{1}{\sqrt{2 \pi}} \sum_{n \in \mathbb{Z}} \re^{- \ri \alpha n} \rv_n~~, 
\label{newc1}
\end{equation}   
where $\alpha \in [ - \pi, \pi]$ is an angle variable, or 
$ \ru'_k \equiv \sqrt{\ell} \ru'_\alpha$,   $ \rv'_k \equiv \sqrt{\ell} \rv'_\alpha$, with 
$k=\alpha / \ell $, $(\ru'_k)^* = \ru'_{-k}$, $(\rv'_k)^* =  \rv'_{-k}$,  $\{\ru'_k, \rv'_{k'} \} = \delta(k-k')$. Because 
\begin{equation}   
\ru_{n \ell} = \frac{1}{\sqrt{2 \pi}} \intk  ~\re^{- \ri k \ell n} \ru'_k~~,~~
v_{n \ell} = \frac{1}{\sqrt{2 \pi}} \intk  ~\re^{ \ri k \ell n} \rv'_k ~~,\label{newc2}
\end{equation}   
for the interaction term  in (\ref{ham1}) one obtains
$$
\ell \sum_{n \in \mathbb{Z}}  \frac{m \kappa}{2} (\ru_n- \ru_{n-1})^2 = 2m \kappa 
\intk \sin^2( \frac{k \ell}{2} ) \ru'_k \ru'_{-k} ~~,  
$$
so that 
\begin{equation}   
 H= \intk ~H'_k~~,~~ H'_k =  \frac{\rv'_k \rv'_{-k}}{2m} + \frac{m \omega_k^2}{2} 
\ru_k' \ru'_{-k}~~,
\end{equation}   
where $\omega_k^2 = \omega_0^2 + 4 \kappa \sin^2 (k \ell /2) = \omega_0^2 + v^2_\ell k^2 j_0^2  (k \ell /2)$, $v_\ell = \ell \sqrt{\kappa}$. The equations of motion 
\begin{equation}
\dot{\ru}'_k = \rv'_{-k} /m~~,~~ \dot{\rv}'_k = -m \omega_k^2 \ru'_{-k}~~, \label{duvp} 
\end{equation}
show that there is a "pure" oscillator mode at $k=0$, while for $k \ne 0$ one can find 
both oscillator modes, such as $\ru'_k(t)= \cos \omega_k t ~ {\ru'_k}^0$, 
$({\ru'_k}^0)^*= {\ru'_-}^0_k$, and "plane rotator" modes \cite{qdpr}, 
${\ru'_k}^\pm(t) = \re^{ \pm \ri \sigma_k \omega_k t} {\ru'_k}^0$, $\sigma_k = k / \vert k
\vert$, of constant  $\vert {\ru'_k}^\pm(t) \vert$. At $\omega_0=0$, or intermediate 
wavelengths $\lambda = 2 \pi /k$, $ 2 \ell \ll \lambda \ll 2 \pi v_\ell / \omega_0$, such that 
$\omega_k \approx v_\ell \vert k \vert$, the "rotational" amplitudes ${\ru'_k}^\pm (t)$  provide
by (\ref{newc2}) 
\begin{equation}
\ru_{n \ell}^\pm (t)  = \frac{1}{\sqrt{2 \pi}} \intk  ~\re^{- \ri k (\ell n \mp v_\ell t)} 
{\ru'_k}^0~~, \label{rsol}
\end{equation}
or, in the continuous limit, $\ru^\pm(x,t) = \ru^0(x \mp v t)$. In this case, the initial
conditions $\ru^0, \dot{\ru}^0$ in (\ref{spwe}) are not independent anymore, but related by
$(\dot{\ru}^0)^\pm = \mp v {\mathfrak d} \ru^0$, ${\mathfrak d} \ru^0 (x) \equiv d \ru^0/dx$,  such that 
\begin{equation}
\ru^\pm(t) = \mp v \mathcal{E}_t  \tilde{*} {\mathfrak d} \ru^0 + 
\partial_t \mathcal{E}_t \tilde{*} \ru^0 \equiv \mathcal{G}^\pm \tilde{*} \ru^0~~,  \label{upm}
\end{equation}
where $\mathcal{G}^\pm = (\partial_t \mp v \partial_x) \mathcal{E}$. Because $\mathcal{G}^\pm_t$  is the fundamental solution of the transport equation 
$(\partial_t \pm v \partial_x) \mathcal{G}^\pm_t = \delta(x) \delta(t)$, linear in 
$\partial_t$, the functions $\ru^\pm(t)$ may also be called  phonon wave modes. 
\\ \indent
For the further study, a more suitable set of coordinates is provided by the functionals $\psi'_k, {\psi'_k}^*$  of $\ru,\rv$,
\begin{equation} 
 \psi'_k = \sqrt{\frac{m \omega_k}{2}} ( {\ru'_k}^* + \frac{ \ri \rv'_k}{m \omega_k} )~~,~~
\{ {\psi'_k}^* , \psi'_{k'} \} = \ri \delta (k-k')~~, \label{psik}
\end{equation}
adapted to the solutions $\ru^\pm$ as ${\psi'_k}^\pm \equiv {\psi'_k}\vert_{\ru^\pm}=  \sqrt{m \omega_k /2} (1 \pm \sigma_k) ({\ru'_k}^\pm)^* $ is independent of $\rv$. In these variables $H'_k = \omega_k (\vert \psi'_k \vert^2 + \vert \psi'_{-k} \vert^2)/2$, such that $ \ri \dot{\psi}'_k = \omega_k \psi'_k$. The relationship  $\psi'_k = \sqrt{\eta_k} \re^{\ri \varphi_k}$ 
to the "harmonic oscillator" action-angle variables $(\eta_k, \varphi_k)$ shows that during time evolution $\eta_k = \vert \psi'_k \vert^2$ is  a constant, while $\varphi_k (t) = - \omega_k t + \varphi^0_k$. In terms of $\psi'_k, {\psi'_k}^*$ the canonical symplectic form $\Omega_M$ on $M$,
\begin{equation}
 \Omega_M = \sum_{n \in \mathbb{Z}} d\ru_n \wedge d\rv_n = \ell         
 \sum_{n \in \nz} d\ru_{n \ell} \wedge d\rv_{n \ell} ~~,
\end{equation}
becomes 
\begin{equation}
\Omega_M = \ri \intk~ d \psi'_k \wedge d{\psi'_k}^* = \intk~ d \eta_k \wedge d \varphi_k~~,
\end{equation}
so that $A_D =\vert \int_D \Omega_M \vert =2 \pi \intk ~\eta_k$ is the  phase-space area 
of the invariant subset $D \subset M$, bounded by the "wave-orbit" cylinder $\partial D =\{ (\varphi_k \in [ - \pi, \pi],  \eta_k ={\rm constant}), k \in [- \pi/\ell,  \pi/\ell] \}$.  \\ \indent
The coordinates $\psi'_k$ can also be represented by functions $\psi_{n \ell}$ on $\ell \nz$, 
\begin{equation}
\psi_{n \ell} = \frac{1}{\sqrt{2 \pi}}\intk ~\re^{\ri k \ell n} \psi'_k~~,~~ 
\langle \psi \vert \psi \rangle \equiv \ell \sum_{n \in \nz} \psi_{n \ell}^* 
\psi_{n \ell} = \frac{A_D}{ 2 \pi} ~~. \label{psinl}        
\end{equation}
In this representation $H = \ell^2 \sum_{n,n' \in \nz} \psi^*_{n \ell} \hat{\mathcal H}_{nn'} \psi_{n' \ell} \equiv  \langle \psi \vert \hat{\mathcal H} \vert \psi \rangle$, with
\begin{equation}
\hat{\mathcal H}_{nn'} =\frac{1}{2 \pi} \intk ~ \omega_k \re^{\ri k \ell (n-n')}~~, 
\end{equation}   
and $\ri \partial_t \psi = \hat{\mathcal H} \psi$.  Because $A_D$ in (\ref{psinl}) is a positive scalar, a constraint $A_D= N \hp$, $N \in \mathbb{N}$, to integer multiples of the Planck's constant $\hp$ \cite{planck}, allows the interpretation of $\psi / \sqrt{\hbar}$ as amplitude for linear density/probability\footnote{The normalization of the 1-phonon (RPA) excited states, by "action integrals" quantization, was noticed before in the case of the constrained quantum dynamics on coherent states manifolds \cite{qdss, cev}.} distributions of $N$ action quanta $\hp$ (phonons)  along the coordinate $x= n \ell \in \ell \nz$. However, a phonon momentum $p$, conjugate to $x$, cannot be defined like in the case of the inertial motion, because when $\ell \nz$ is taken as a configuration space, the tangent space does not exist\footnote{A similar situation occurs in the more general case of discrete fractal sets \cite{dya}.}.  An alternative is to define $p \equiv \hbar k$, where $k$ is the Fourier dual variable to $x$, known to appear in linear combinations with the
mechanical momentum\footnote{When $p = \hbar k$, the scaled canonical 1-form $pdx/\hp$   represents the number of wavelengths $\lambda$ in the interval $dx$.}  at the massive particles \cite{qdpr}. For consistency, the obtained phase-space $M_\ell = \ell \nz \times \hbar [- \pi/\ell, \pi/\ell]$ carries indeed an intrinsic "granular structure", with $\hp = \ell \cdot 2 \pi \hbar / \ell$, as the $\ell$ - independent area of an "elementary cell", until the continuous limit, $M_\ell  \rightarrow \mathbb{R}^2$. A phase-space distribution function ${\sf f}_N$ of the $N$ phonons  can be further defined  by the Wigner transform of $\psi / \sqrt{\hbar}$,
\begin{equation}     
{\sf f}_N (x,p) = \frac{1}{2 \pi \hbar} \int dk'~\re^{- \ri k' p} \psi_{x+  \frac{\hbar k'}{2}}
\psi^*_{x- \frac{\hbar k'}{2}} = \frac{1}{2 \pi \hbar^2} \int dk~\re^{\ri kx} \psi'_{
\frac{p}{\hbar}+ \frac{k}{2}} (\psi')^*_{\frac{p}{\hbar}- \frac{k}{2}}~~,~~ 
\end{equation}
such that $ \int dxdp~{\sf f}_N (x,p) = A_D/ \hp= N$. Considering  here $\psi'_k (t) = \sqrt{\eta_k} \re^{- \ri \omega_k t}$, ($\varphi^0_k=0$), and $\omega_a-\omega_b \approx (a-b) v_{\rm g}$, where $v_{\rm g} = d \omega_k/dk \vert_{k_p}$ denotes the group velocity at $k_p = (a+b)/2 = p/ \hbar$, ${\sf f}_N$ becomes
\begin{equation}     
{\sf f}_N (x,p,t) = \frac{1}{2 \pi \hbar^2} \int dk~\re^{\ri k(x -v_g t)} \sqrt{\eta_{
\frac{p}{\hbar}+ \frac{k}{2}} \eta_{\frac{p}{\hbar}- \frac{k}{2}} }~~. 
\end{equation}
In the case of a Gaussian function $\eta_k^G= C_N \sqrt{g/ \pi } ~\re^{ - g (k-k_0)^2}$,
parameterized by $k_0$, $g$, and  $C_N = A_D/2 \pi = N \hbar$,  one obtains 
\begin{equation}
{\sf f}_N^G (x,p,t) = \frac{N}{\pi \hbar} \re^{ - g (p- \hbar k_0)^2/ \hbar^2
- (x - v_{\rm g} t)^2/g} ~~, 
\end{equation}
corresponding to an ensemble of $N$ phonons having the average momentum $p_0=\hbar k_0$, moving coherently along the $x$-axis with the velocity $v_{\rm g}$ (e.g. as observed in  \cite{fsdi}). \\ \indent
Similarly to $\psi$ of (\ref{psinl}), one can define a function  $\Phi$, 
\begin{equation}
\Phi_{n \ell} = \frac{1}{\sqrt{2 \pi} } \intk ~\re^{\ri k \ell n} \Phi'_k~~,~~ \Phi'_k \equiv  \sqrt{\omega_k} \psi'_k = \frac{\ri}{\sqrt{2m}} (\rv'_k + \frac{\ri  \dot{\rv}'_k}{\omega_k} )~~,      
\end{equation}
and in the limit  $M_\ell  \rightarrow \mathbb{R}^2$,  the quasi-energy density ${\sf f}_E$, 
\begin{equation}     
{\sf f}_E (x,p) = \frac{1}{2 \pi } \int dk'~\re^{- \ri k' p} \Phi_{x+  \frac{\hbar k'}{2}}
\Phi^*_{x- \frac{\hbar k'}{2}} ~~,~~H \equiv \int dxdp~{\sf f}_E (x,p) =E ~~,
\end{equation}
resembling the photon phase-space distribution presented in \cite{bb}.

\section{Electromagnetic action waves and photon distributions}
On the space-time manifold $\mathbb{R}^4$, with coordinates $(x_0,{\bf x} )$, $x_0=ct$, ${\bf x} = (x_1,x_2,x_3)$,  and metric $ds^2 =- dx_0^2+ d{\bf x}^2$, the electromagnetic field ${\bf E}, {\bf B}$ provides a 2-form  $\omega_f \in \wedge^2(\mathbb{R}^4)$,
\begin{equation}
\omega_f = {\bf E} \cdot d {\bf S}_0  - {\bf B} \cdot d {\bf S}~~,
\end{equation}
where $d {\bf S}_0 \equiv dx_0 \wedge d{\bf x}$, $ d{\bf S} \equiv ( dx_2 \wedge dx_3, - dx_1 \wedge dx_3, dx_1 \wedge dx_2)$. This form has the Hodge  dual\footnote{In $\mathbb{R}^n$, with the volume element $d^n x\equiv dx_1 \wedge...\wedge  dx_n$, and metric $ds^2 \equiv \sigma_1 dx_1^2+...+\sigma_n dx_n^2$, $\sigma_i = \pm 1$, if  $d^nx = \alpha \wedge \beta$, where $\alpha = dx_{i_1} \wedge ...\wedge dx_{i_p}$, $p \le n$,  then $* \alpha \equiv \sigma_{i_1} ...\sigma_{i_p} \beta$.}
\begin{equation}
*\omega_f = -{\bf B} \cdot d {\bf S}_0 - {\bf E} \cdot d {\bf S}~~,
\end{equation}
and the Maxwell equations can be derived from the "intrinsic" expressions 
\begin{equation}
 d \omega_f = J_m~~,~~d* \omega_f = J_e~~, 
\end{equation}
(with $\hat{*}$-duality  \cite{fdb} and $\hat{P}_c$  $\iota$-symmetry\footnote{$\iota \equiv 
[\hat{*} ; \hat{P}_c]$, $\iota^2 = [-1 ; -1]$, defined by $\hat{*}(\omega_f;*\omega_f) \equiv (*\omega_f; - \omega_f)$, $\hat{P}_c (J_m;J_e)  \equiv(J_e; -J_m)$.}), where $J= {\bf j} \cdot dx_0 \wedge d{\bf S}/c  - \rho d^3x$, are the current 3-forms for magnetic ($m$), respectively electric ($e$), charges. \\ \indent
Due to the absence of magnetic monopoles in our 3d space, $J_m=0$, and the first equation $d \omega_f =0$, (or $ \nabla \cdot {\bf B} =0$, $\nabla \times {\bf E} = - \partial_0 {\bf B}$, $\partial_0 \equiv \partial / \partial x_0$), ensures the existence of a potential 1-form $\theta_f =      
A_0 dx_0+{\bf A} \cdot d{\bf x}$, ($-A_0=V$ is the Coulomb potential), defined up to a
"gauge" term $df=\partial_0 f dx_0+ \nabla f \cdot d{\bf x}$, such that $\omega_f = - 
d \theta_f$, (${\bf E} = - \partial_0 {\bf A} + \nabla A_0$, ${\bf B} = \nabla \times 
{\bf A}$). In the following, the gauge (the inital conditions), will be fixed to have $\nabla \cdot {\bf A} =0$.
\\ \indent
The potential ${\mathfrak A}=(A_0,{\bf A})$  provides a more restricted set of "coordinate" variables for the field, because the second equation $d* \omega_f = J_e$ can be obtained from the variational principle $\delta_{\mathfrak A} (S_f+S_{int})=0$, where $\delta_{\mathfrak A}$ denotes the functional variation with respect to ${\mathfrak A}$, $S_f = - (\omega_f, \omega_f)_{ip}/ 2c$,  $S_{int} = - (\theta_f, *J_e)_{ip}/c$, and $( \alpha, \beta)_{ip} \equiv \int_{\mathbb{R}^4} \alpha \wedge * \beta$  is the internal product. \\ \indent
The current $J_e$ can be decomposed as $J_e = J_f+ d \omega_{pol}$, where $J_f$ is due to the free charges, and $\omega_{pol} = - {\bf M} \cdot d {\bf S}_0 + {\bf P} \cdot d {\bf S}$ 
is the polarization form, with ${\bf P} = \chi_e {\bf E} + {\bf P}_0$,   
${\bf M} = \chi_m {\bf H} + {\bf M}_0$, the polarization, respectively magnetization vectors,
having both induced ($\sim \chi$), and intrinsic\footnote{A term  ${\bf P}_0/{\bf M}_0$ can stand for an antenna-source of electric/magnetic-type Hertz vector waves. In \cite{stan}, the intrinsic atomic electric and magnetic dipole moments are parts of a photon detection device.} terms. Thus, the equation  $d*\omega_f = J_e$ becomes $d(*\omega_f - \omega_{pol}) = J_f$, and further, in a homogeneous, isotropic medium, with no intrinsic polarization (${\bf P}_0=0$, ${\bf M}_0=0$), it provides 
\begin{equation}
\nabla \cdot {\bf D} = \rho_f~~,~~ \nabla \times {\bf H} = \partial_0 {\bf D} + {\bf j}_f /c ~~,
\end{equation} 
where ${\bf D} = \epsilon {\bf E}$,  ${\bf H} = {\bf B} / \mu$, and $\epsilon = 1+ \chi_e$, 
$\mu = 1+\chi_m$ (usually denoted   $\epsilon_r, \mu_r$). 
By introducing the complex vector ${\bf F} = \sqrt{\epsilon} {\bf E} + \ri \sqrt{\mu} 
{\bf H}$ \cite{ll}, the time derivatives  $\partial_t {\bf E}$, $ \partial_t {\bf H}$ can also be expressed in the form 
\begin{equation}
\partial_t {\bf F} = - \ri v \nabla \times {\bf F} - {\bf j}_f / \sqrt{\epsilon}
~~,~~ v = c/ \sqrt{\epsilon \mu}~~.  \label{dtf}
\end{equation}
The related energy flow is described by the transport equation
\begin{equation}
\partial_t w +\nabla \cdot {\bf Y} + {\bf E} \cdot {\bf j}_f=0 ~~, \label{efc}
\end{equation}
where $w = ({\bf E} \cdot {\bf D} +{\bf H} \cdot {\bf B})/2 = {\bf F} \cdot {\bf F}^*/2$, 
($=T_{00}$), is the energy density, and ${\bf Y}=c  {\bf E} \times {\bf H}= \ri v  
{\bf F} \times {\bf F}^*/2$, is the Poynting vector.  \\ \indent
The linearity in $\partial_t$ of  (\ref{dtf}) was a reason in \cite{bb} to interpret ${\bf F}$ 
as quantum photon wave function, while the linearity of (\ref{efc}) has been used in \cite{em2f} to find a classical Hamilton function for photons. Unlike ${\bf F}$ and $w$, the potential ${\bf A}\vert_{\nabla \cdot {\bf A} =0} $ satisfies the inhomogeneous wave equation
\begin{equation}
\partial_t^2 {\bf A} - v^2 \Delta {\bf A}=  c  {\bf j}_f / \epsilon - 
c \partial_t \nabla V  ~~, \label{ewe} 
\end{equation}
containing $\partial_t^2$. 
In the absence of sources (${\bf j}_f=0$, $\partial_t \nabla V =0$), this equation can be 
obtained by using the total field energy $E = \int d^3x ~w$ as a Hamiltonian functional
\begin{equation}
H_F = \frac{1}{2} \int d^3x ~[ \mu v^2 {\bf \Pi}^2 + \frac{1}{\mu} 
(\nabla \times {\bf A})^2] ~~,
\end{equation}
of the canonical coordinates ${\bf A}$, ${\bf \Pi} = \dot{\bf A}/ \mu v^2$,    
$\{ {A}_{i{\bf x}},\Pi_{j {\bf x'}} \} =\delta_{ij} \delta^3( {\bf x} - {\bf x}')$, ($\nabla 
\cdot {\bf A} =0$  is only a constraint on the solutions). Similarly to  (\ref{newc1}), the Fourier transform  of ${\bf A}_{\bf x}$ introduces the complex coordinates ${\bf A'}_{\bf k}$,  ${\bf A'}_{\bf k}^*= {\bf A'}_{-\bf k}$, such that  
\begin{equation}
{\bf A}_{\bf x} = \frac{1}{(2 \pi)^{3/2}} \int d^3k ~ \re^{ - \ri {\bf k} \cdot {\bf x}}
{\bf A'}_{\bf k}~~,~~ {\bf k} =(k_1,k_2,k_3)~~,  \label{vep}
\end{equation} 
and if  $\partial_t^2 {\bf A} = v^2 \Delta {\bf A}$, $\nabla \cdot {\bf A}=0$, then 
$\ddot{\bf A'}_{\bf k}= - \omega_k^2  {\bf A'}_{\bf k}$,  $\omega_k = v 
\vert {\bf k} \vert$, ${\bf k} \cdot {\bf A'}_{\bf k}=0$.  The previous solutions
(\ref{upm}) correspond in this case to photon wave modes, having "rotational" amplitudes 
$\vert {\bf A'}_{\bf k} \vert=$ constant, such as the linearly polarized plane 
waves $ {\bf A}^\pm_{lp} (x_1,t)= \ru^0 (x_1 \mp vt) {\bf e}_{lp}$, where $\ru^0$ is the 
initial condition in (\ref{upm}), and ${\bf e}_{lp} =(0,e_2,e_3)$ is a constant, real  polarization vector. More specific solutions, with circular polarization, are $ {\bf A}^\pm_{cp}  (x_3,t)= A_\perp 
(\cos \phi_\pm, \pm \sin \phi_\pm, 0)$,  $\phi_\pm =  k( x_3 \mp vt) $, related to the 
real  part of the complex vector potential ${\bf U}$  for ${\bf F}$,  (Appendix 1), by  $ {\bf A}^\pm_{cp} = \sqrt{\mu}({\bf U}_{\pm 1} + {\bf U}_{\pm 1}^*)/2$, where
\begin{equation}
 {\bf U}_\sigma (x_3,t)= \sqrt{\frac{2}{\mu}} A_\perp  \re^{- \ri  k( x_3- \sigma vt)} {\bf e}_{p \sigma}~~,~~{\bf e}_{p \sigma} =  \frac{1}{\sqrt{2}}(1, \ri \sigma,0)~~,~~
\sigma = \pm 1~~. 
\end{equation}
 Aside the coordinates ${\bf A'}_{\bf k}$, one can define variables similar to (\ref{psik}),  
\begin{equation}
 \vec{\psi' }_{\bf k} = \sqrt{ \frac{\omega_k}{2 \mu v^2}} ({\bf A'}_{\bf k}^* +
\frac{\ri}{\omega_k} \dot{\bf A'}^*_{\bf k} ) \equiv
 \sum_{\tau=1,2} \sqrt{\eta_{\bf k \tau}} \re^{ \ri \varphi_{\bf k \tau}} 
{\bf e}_{\bf k \tau}~~, \label{phwfk}
\end{equation}
where ${\bf e}_{\bf k \tau}$ are two real polarization vectors, 
${\bf e}_{\bf k \tau} \cdot {\bf e}_{\bf k \tau'} = \delta_{\tau \tau'}$,
orthogonal to ${\bf k}$, and $\eta_{\bf k \tau}, \varphi_{\bf k \tau}$ are the 
"oscillator"action-angle coordinates. Because $\ri \partial_t \vec{\psi' }_{\bf k} = \omega_k 
\vec{\psi' }_{\bf k}$,  the associated "action field" 
\begin{equation}
\vec{\psi}_{\bf x} = \frac{1}{(2 \pi)^{3/2}} \int d^3k ~ \re^{  \ri {\bf k} \cdot {\bf x}}
\sqrt{ \frac{\omega_k}{2 \mu v^2}} ({\bf A'}_{\bf k}^* +
\frac{\ri}{\omega_k} \dot{\bf A'}^*_{\bf k} ) ~~ \label{phwf}   
\end{equation} 
satisfies $ \ri \partial_t \vec{\psi} = \hat{\mathcal H}_f \vec{\psi}$, with       
\begin{equation}
\hat{\mathcal H}_{f{\bf xx}'}=\frac{1}{(2 \pi)^3} \int d^3k ~\omega_k  
\re^{  \ri {\bf k} \cdot ({\bf x}- {\bf x}')}~~,~~\hat{\mathcal H}^2_f=-v^2 \Delta~~.
\end{equation}
Expressing the amplitudes ${\bf A'}_{\bf k}$ of (\ref{vep}) in terms of $\vec{\psi}$, one
obtains
\begin{equation}
{\bf A}_{\bf x} = \frac{c}{(2 \pi)^{3/2}} \int \frac{d^3k}{\sqrt{2 \epsilon \omega_k}} ~ 
(\re^{ \ri {\bf k} \cdot {\bf x}} \vec{\psi'}_{\bf k}+\re^{ -
\ri {\bf k} \cdot {\bf x}} \vec{\psi'}_{\bf k}^*)~~. \label{vep1}
\end{equation} 
If we "quantize" ${\bf A}_{\bf x}$ by replacing for $\sqrt{\eta_{\bf k \tau}} \re^{ \ri 
\varphi_{\bf k \tau} }$ and $\sqrt{\eta_{\bf k \tau}} \re^{- \ri 
\varphi_{\bf k \tau} }$ in $\vec{\psi'}_{\bf k}$, $ \vec{\psi'}_{\bf k}^*$  the operators $\sqrt{\hbar} \hat{a}_{\bf k  \tau}$, respectively $\sqrt{\hbar} \hat{a}_{\bf k  \tau}^\dagger$,  with (scaled) commutator $[\hat{a}_{\bf k \tau}, \hat{a}_{\bf k' \tau'}^\dagger] = \delta_{\tau \tau'} \delta^3({\bf k}-{\bf k}')$, then in vacuum ($\epsilon =1$), using the units $c=1$, $\hbar=1$,  (\ref{vep1}) becomes the field operator of \cite{bd}, p. 90. However, one can proceed also by considering  (\ref{phwf}), after a suitable normalization, as photon wave function. 
\\ \indent   
The functional 2-form 
$ \Omega_F = \int d^3x ~d {\bf A} \cdot \wedge d {\bf \Pi} 
= \ri \int d^3x ~d \vec{\psi} \cdot \wedge d \vec{\psi}^* $
indicates that $A_W = 2 \pi \langle \vec{\psi}  \vert \vec{\psi} \rangle=  2 \pi   \int d^3 k( \eta_{{\bf k}1}+\eta_{{\bf k}2}) $ represents the total  phase-space area bounded by a wave-like orbit $\eta_{\bf k \tau}=$ constant, such that the quantization condition $A_W=N \hp$, $N \in \mathbb{N}$, can be used to normalize $\vec{\psi}$ to a given number of action quanta (photons). Moreover, a photon quasi-density ${\sf f}_N$ on the 
phase-space $T^* \mathbb{R}^3$ can be obtained by the Wigner transform of $\vec{\psi}/ 
\sqrt{\hbar}$,   
\begin{equation}
{\sf f}_N ({\bf x}, {\bf p}) = \frac{1}{(2 \pi)^3 \hbar} \int d^3k' ~  
\re^{ - \ri {\bf k}' \cdot {\bf p}}    ~~
\vec{\psi}_{{\bf x}+  \frac{\hbar {\bf k}'}{2}} \cdot \vec{\psi}^*_{{\bf x}- 
\frac{\hbar {\bf k}'}{2}} ~~, \label{phqd}
\end{equation}
$\int d^3x d^3p ~{\sf f}_N ({\bf x}, {\bf p}) =N$.  According to the general identity
\cite{qdpr, qdq} 
\begin{equation}
\int d^3x d^3p~ f_1({\bf x},{\bf p}) f_2 ({\bf x},{\bf p})   \equiv \hp^3 Tr( \hat{f}_1 \hat{f}_2)~~, 
\end{equation}
 the total wave-field  energy  $E_W = \int d^3x~w = \langle \vec{\psi} \vert  \hat{\mathcal H}_f \vert \vec{\psi} \rangle = \hp^3 Tr( \hat{\sf f}_N \hbar \hat{\mathcal H}_f )$ becomes in the phase-space representation $E_W= \int d^3x d^3p ~{\sf f}_N ({\bf x}, {\bf p}) \epsilon_p $,
where $\epsilon_p =  \hbar \omega_{p/ \hbar}= v \vert {\bf p} \vert$ is the photon energy. 
At a finite temperature $T$, the equilibrium energy density takes the form  $\tilde{w}_T = \int d^3p ~{\sf f}^T  ({\bf x}, {\bf p}) \epsilon_p$, independent of $N$, where 
\begin{equation}
{\sf f}^T ({\bf x},{\bf p}) = \frac{1}{\hp^3} \frac{2}{\re^{\epsilon_p /\kb T} -1} \label{pld} 
\end{equation}
is the Planck distribution function \cite{planck} (Appendix 2). The Wien's displacement law in vacuum, written as $\lambda_m p_T = 1.26 \hbar$, indicates that for a given "thermal momentum" $p_T = \kb T/c$, the maximum of the spectral energy density $U_\lambda^T$, ($\tilde{w}_T= \int d \lambda~ U_\lambda^T$), appears at a wavelength $\lambda_m$ close to the "dual" value $\hp /p_T$.  The number of thermal photons in a cube of volume $\lambda_m^3$  is 
 $N_m = \lambda_m^3 \int d^3p ~{\sf f}^T ({\bf x},{\bf p})=0.48$, at any temperature. 

\section{Concluding remarks} 
The fundamental constituents of matter can be associated, by structural stability, with  elementary adiabatic invariants, abstract entities encountered in the treatment of complex dynamical systems, including classical fields. The "wave quanta"  are such invariants, as phase-space area elements $\hp = 2 \pi \hbar= 4.1$ meV$\cdot$ps, defining a partition of the total action integral for the wave-field.
\\ \indent
In the example of Section 2, (\ref{psik}) and  (\ref{psinl}) define the  functional $\psi_{[\ru], [\rv]}(x,t)$,  normalized (using $A_D=N \hp$) by summation over $x$ at a fixed wave configuration $[\ru],[\rv]$, instead of integrating over all configurations at given $x$. Therefore, if $N$ is large, it should be regarded as density amplitude for a quasi-classical  ensemble of $N$ action quanta $\hp$. Nevertheless, it becomes closer to a  true Schr\"odinger wave function, depending only on the "field coordinates"  $\ru$, on the subset of the "rotational" solutions $\ru^\pm$, when it reduces to $\psi_{[\ru]}^\pm(x,t)$. In particular, for $N=1$, $\vert \psi_{[\ru]}^\pm(x,t)  \vert^2/ \hbar$ can  be interpreted as quantum  probability density for the spatial localization of a phonon.   \\ \indent
The electromagnetic field is related to area elements ($d {\bf S}_0=*d {\bf S}$)  in space-time ($\mathbb{R}^4$), and a suitable definition of its canonical coordinates  is more difficult to find.  Thus, in the perspective of the extended dynamics on the phase-space $T^*\mathbb{R}^4$  \cite{rpw, em2f}, the "equations of motion" $d*\omega_f =J_e$, obtained from the variational principle $\delta_{\mathfrak A} (S_f+S_{int})=0$, are essentially "static". For a free field, generated initially by $J_f$, described by the variables ${\bf A},{\bf \Pi}$, one can define the action  wave function  $\vec{\psi}$ of (\ref{phwf}), and the photon phase-space distribution (\ref{phqd}), in agreement both with standard QED, and the classical results on thermal radiation. However,  one can find also free fields described by the complex potential ${\bf U}$ (Appendix 1), resembling more the massive quantum particles, by not only being unrelated to a classical source, but "collapsing" to ${\bf A}= \sqrt{\mu}({\bf U} + {\bf U}^*)/2$ in the space regions where $J_f \ne 0$. Such fields could be generated like the electrons in $\beta$ decay, by "action at a distance" from the source, and may have unusual, longitudinal components. \\

\noindent
{\bf Appendix 1: The complex electromagnetic vector potential } \\
The complex field  ${\bf F} = \sqrt{\epsilon} {\bf E} + \ri \sqrt{\mu} 
{\bf H}$ is associated to the complex 2-form $\tilde{\omega}_f={\bf F} \cdot d {\bf Z}$,    
\begin{equation}
\tilde{\omega}_f = \frac{1}{\sqrt{ \mu}} \omega_f   
-\frac{\ri}{\sqrt{ \epsilon}} (*\omega_f - \omega_{pol}) ~~,~~
d{\bf Z} = \frac{1}{\sqrt{\epsilon \mu}} d {\bf S}_0  + \ri d {\bf S}~~.
\end{equation}
In the absence of free charges and currents $d\tilde{\omega}_f=0$, and there exists\footnote{The  equation $i_\zeta \tilde{\omega}_f=0$ for the real characteristic vector $\zeta= l_0 \partial_0+{\bf l}\cdot \nabla$, requires ${\bf l}\cdot {\bf F}=0$ and  $\vert {\bf l} \vert =v \vert l_0 \vert/ c$ \cite{em2f}, similar to $\vert {\bf Y} \vert = v  w $ and $ \vert {\bf p}_A \vert =v  \hbar \omega / c^2$, where ${\bf p}_A$ in the Abraham's photon momentum.}  a 
complex 1-form $\tilde{\theta}_f =U_0 dx_0+{\bf U} \cdot d {\bf x}$, (defined up to a 
gauge term), where ${\mathfrak U} =(U_0,{\bf U})$ is a complex potential, such that 
$\tilde{\omega}_f= - d \tilde{\theta}_f $.  This equality reduces to ${\bf F} = \sqrt{\epsilon \mu} ( - \partial_0 {\bf U} + \nabla U_0 ) = \ri \nabla \times {\bf U}$, so that ${\bf U}$ should satisfy  
\begin{equation}
\ri (\partial_t {\bf U} -c \nabla U_0 ) = v \nabla \times {\bf U}~~,~~~v= 
c/ \sqrt{\epsilon \mu} ~~.
\end{equation}
By choosing the gauge such that $U_0=0$, one obtains $\ri \partial_t {\bf U}  = v 
\nabla \times {\bf U}$, (of the form  (\ref{dtf}) at ${\bf j}_f=0$), which means that in the 
Fourier expansion 
\begin{equation}
{\bf U}({\bf x},t) = \frac{1}{(2 \pi)^{3/2}} \int d^3k ~ \re^{ - \ri {\bf k} \cdot {\bf x}}
{\bf U'}_{\bf k}(t)~~,  
\end{equation} 
the vector ${\bf U'}_{\bf k}$ performs a precession around ${\bf k}$, according to 
 $\dot{\bf U}'_{\bf k}= - v {\bf k} \times {\bf U'}_{\bf k}$. For the complex vectors  ${\bf U'}_{\bf k \sigma}$ such that ${\bf k} \times {\bf U'}_{\bf k \sigma} = - \ri \sigma  \vert {\bf k} \vert {\bf U'}_{\bf k \sigma} $,  where $\sigma = \pm 1$ is the helicity, this precession reduces to a phase factor, and ${\bf U'}_{\bf k \sigma}(t) =  \re^{ \ri \sigma \omega_k t}{\bf U'}_{\bf k \sigma} (0)$, $\omega_k = v \vert {\bf k} \vert$, similar to $\vec{\psi'}_{\bf k} (t)= \re^{- \ri \omega_k t} \vec{\psi'}_{\bf k}(0)$ in  (\ref{vep1}). Another interesting solution is 
\begin{equation} 
{\bf U}(\rho, \varphi, z,t) = \re^{\ri k(z - v t)} (\rho {\bf e}_p 
+ \ri \frac{\sqrt{2}}{k} {\bf e}_z )  ~~,~~ \nabla  \cdot {\bf U}=0~~,
\end{equation}
where $(\rho, \varphi, z)$ are the cylindrical coordinates with unit tangent vectors
$({\bf e}_\rho, {\bf e}_\varphi, {\bf e}_z)$, and ${\bf e}_p = ({\bf e}_\rho+ \ri 
{\bf e}_\varphi)/ \sqrt{2}$ has definite helicity, ${\bf e}_z \times {\bf e}_p= - 
\ri{\bf e}_p$. This result shows that the cylindrical symmetry, known for its special 
properties in the case of the gravitational waves \cite{gw}, also provides peculiar 
electromagnetic waves, having both transversal and longitudinal components. 
\\

 \noindent
{\bf Appendix 2: The quantum noise and thermal radiation} \\
An ensemble of free photons in a homogeneous, isotropic, non-dispersive medium, can be 
described by the phase-space distribution function (\ref{phqd}), 
$$
{\sf f} ({\bf x}, {\bf p}) = \frac{1}{(2 \pi)^3 \hbar} \int d^3k' ~  
\re^{ - \ri {\bf k}' \cdot {\bf p}}    ~
\vec{\psi}_{{\bf x}+  \frac{\hbar {\bf k}'}{2}} \cdot \vec{\psi}^*_{{\bf x}- 
\frac{\hbar {\bf k}'}{2}} = \frac{1}{(2 \pi)^3 \hbar^4} \int d^3k ~  
\re^{ \ri {\bf k} \cdot {\bf x}}    ~
\vec{\psi'}_{\frac{\bf p}{\hbar}+  \frac{\bf k}{2}} \cdot \vec{\psi'}^*_{
\frac{\bf p}{\hbar}-  \frac{\bf k}{2}} ~~, 
$$
where $\vec{\psi'}_{\bf k} (t)= \re^{- \ri \omega_k t} \vec{\psi'}_{\bf k}(0)$. Presuming
that the main contribution to the integral arises from the subset $\vert {\bf k} \vert \ll
2 \vert {\bf p} \vert / \hbar$, then for ${\bf a}= {\bf p}/\hbar+  {\bf k}/2$, 
${\bf b}= {\bf p}/\hbar-{\bf k}/2$, we get $\omega_a - \omega_b \approx ({\bf a}- {\bf b}) 
\cdot \nabla_{\bf k} \omega \vert_{{\bf k}= {\bf p} / \hbar} = {\bf k} \cdot {\bf v}_{\bf p}$, 
$ {\bf v}_{\bf p} = v {\bf p} / \vert {\bf p} \vert$, and 
\begin{equation}
\partial_t {\sf f}  =  - \nabla \cdot {\bf y}~~,~~ {\bf y} =  {\bf v}_{\bf p} {\sf f}~~.
\label{A21}
\end{equation}
If (\ref{ewe}) contains  ${\bf j}_f \ne 0$,  $\nabla \cdot {\bf j}_f = 0$, then $\ddot{\bf A'}_{\bf k}= - \omega_k^2  {\bf A'}_{\bf k}+ c {\bf j}'_{f {\bf k}}/ \epsilon $, and $\vec{\psi'}_{\bf k}$ of (\ref{phwfk}) satisfies $\ri \partial_t \vec{\psi' }_{\bf k} = \omega_k \vec{\psi' }_{\bf k}- {\bf j'}_{f {\bf k}}^* / \sqrt{2 \epsilon \omega_k}$. In particular, when ${\bf j}_f$ reduces to the (quasi) Ohmic current ${\bf j}_\Omega = \sigma_q  {\bf E}$ generated by the ensemble of photons, specified by  
\begin{equation}
{\bf j}'_{\Omega {\bf k}} = - \frac{\sigma_q}{c} \dot{\bf A}'_{\bf k} = - \ri 
\gamma \sqrt{\frac{\epsilon \omega_k}{2}} (\vec{\psi'}^*_{\bf k} - \vec{\psi'}_{-{\bf k}})~~,~~ \gamma =  \frac{\sigma_q}{\epsilon} > 0~~,
\end{equation}
one obtains 
$$\partial_t (\vec{\psi'}_{\bf a} \cdot \vec{\psi'}^*_{\bf b}) = - \ri 
(\omega_a - \omega_b) \vec{\psi'}_{\bf a} \cdot \vec{\psi'}^*_{\bf b} -
\gamma \vec{\psi'}_{\bf a} \cdot \vec{\psi'}^*_{\bf b} + \frac{\gamma}{2}
\vec{\psi'}_{\bf a} \cdot \vec{\psi'}_{-{\bf b}} + \frac{\gamma}{2}
\vec{\psi'}^*_{- {\bf a}} \cdot \vec{\psi'}^*_{\bf b} ~~.$$
The last two terms include the fast oscillating factors $\re^{ \mp \ri (\omega_a + 
\omega_b)t}$ and their contribution to $\partial_t {\sf f}$ can be neglected, so that  the (quasi) Ohmic environment changes (\ref{A21}) into 
\begin{equation}
\partial_t {\sf f}  =  - \nabla \cdot {\bf y} - \gamma {\sf f}~~. \label{A22}
\end{equation}
This shows that the energy loss described by (\ref{efc}), $\partial_t w = -\nabla 
\cdot {\bf Y} - \gamma w$, is due to the overall decrease of the photon number, while the
energy $\epsilon_p = v \vert {\bf p} \vert $ of each photon is a constant, by contrast to
an ensemble of massive particles, where the number (of particles / degrees of freedom)  is a
constant. \\ \indent
 Aside ${\bf j}_\Omega$,  at finite temperature ${\bf j}_f$ has also a random part of
zero  mean (noise) ${\bf j}_n$,  due to the phonon modes and  thermal fluctuations. The effects of ${\bf j}_n$ can be described by extending (\ref{A22}) to a Fokker-Planck equation, but this requires several assumptions on the correlation function between ${\bf j}_n (t)$ and ${\bf j}_n (t')$. Because the Fourier transform of a random field can also be seen as a peculiar weighted average, particularly important is the choice of the defining representation,  either as the coordinate ({\bf x}), or the momentum ({\bf p} = $\hbar {\bf k}$) space. Let us presume that the correlation function is diagonal in the momentum ("quantum") representation, having the general form 
\begin{equation}          
<<{\bf j}'^*_{n {\bf k}} (t) \cdot {\bf j}'_{n {\bf k}'} (t') >> = 2 \delta(t - t') 
\delta^3({\bf k}-{\bf k}') Q_T (\vert {\bf k} \vert) ~~,
\end{equation}          
where $<<...>>$ denotes the average over the statistical ensemble of the free charges, and $Q_T (\vert {\bf k} \vert)$ still needs to be specified. In this case, (\ref{A22}) is extended by a source term $\sim Q_T(\vert {\bf p} \vert/ \hbar) $, becoming
\begin{equation}          
\partial_t {\sf f} + \nabla \cdot {\bf y} = \gamma( \frac{1}{\hp^3} \frac{Q_T}{\sigma_q 
\epsilon_p} - {\sf f})~~. \label{fpe} 
\end{equation}
At equilibrium, $\partial_t {\sf f}=0$, $\nabla \cdot {\bf y}=0$, and   
${\sf f}^T ({\bf x}, {\bf p})  = Q_T/ \hp^3 \sigma_q \epsilon_p$. 
The "classical" Rayleigh-Jeans formula (consistent with the usual radiation reaction force \cite{ddq}), is provided by $Q^{RJ}_T = 2 \sigma_q  \kb T$, and the "quantum" Wien distribution by $Q^W_T = 2 \sigma_q \epsilon_p \re^{ - \epsilon_p/ \kb T}$, where $\kb =0.086$ meV$/$K is the Boltzmann's constant. The  Planck distribution (\ref{pld}) can also be obtained by introducing the density-dependent correction factor $(1+\hp^3 {\sf f}/2)$ to the $Q^W_T$  source term of (\ref{fpe}), accounting for the stimulated emission in each polarization mode of an elementary cell.  One should note though that for applications to realistic situations of interest now, involving large ($N \sim  10^{11}$), non-equilibrium ensembles of short wavelength photons \cite{sz}, further corrections in both source, and absorption terms, are necessary.

\end{document}